\documentclass[%
mathleft,%
final,%
]{an}
\usepackage{graphicx}
\usepackage{times}
\begin{document}

\Pagespan{1}{}
\Yearpublication{2013}%
\Yearsubmission{2012}%
\Month{1}%
\Volume{334}%
\Issue{1}%
\DOI{This.is/not.aDOI}%

\title{The Stellar Activity -- Rotation Relationship}

\author{Nicholas J. Wright\inst{1}\fnmsep\thanks{Corresponding author. \email{nwright@cfa.harvard.edu}}
\and  Jeremy J. Drake\inst{1} \and Eric E. Mamajek\inst{2} \and Gregory W. Henry\inst{3}}

\titlerunning{The Stellar Activity -- Rotation Relationship}
\authorrunning{Wright et al.}
\institute{Harvard-Smithsonian Center for Astrophysics, 60 Garden Street, Cambridge, MA~02138, U.S.A.
\and
Department of Physics and Astronomy, University of Rochester, Rochester, NY~14627, U.S.A.
\and
Center of Excellence in Information Systems, Tennessee State University, Nashville, TN~37209, U.S.A.}

\received{XXXX}
\accepted{XXXX}
\publonline{XXXX}

\keywords{stars: activity, X-rays: stars, stars: late-type, stars: coronae, stars: magnetic fields, stars: rotation}

\abstract{Using a new catalog of 824 solar and late-type stars with X-ray luminosities and rotation periods we have studied the relationship between rotation and stellar activity. From an unbiased subset of this sample the power law slope of the unsaturated regime, $L_X/L_{bol}\propto Ro^\beta$, is fit as $\beta=-2.70\pm0.13$. This is inconsistent with the canonical $\beta=-2$ slope to a confidence of 5$\sigma$ and argues for an interface-type dynamo. Super-saturation is observed for the fastest rotators in our sample and its parametric dependencies are explored. Significant correlations are found with both the corotation radius and the excess polar updraft, the latter theory being supported by other observations. We also present a new X-ray population synthesis model of the mature stellar component of our Galaxy and use it to reproduce deep observations of a high Galactic latitude field. The model, XStar, can be used to test models of stellar spin-down and dynamo decay, as well as for estimating stellar X-ray contamination rates for non-stellar studies.}

\maketitle

\section{Introduction}

Stars across the Hertzsprung-Russell diagram are known to emit X-rays with only a few exceptions. Solar- and late-type stars are thought to generate their X-rays from a magnetically confined plasma known as a corona (Vaiana et al. 1981), which is believed to be driven by the stellar magnetic dynamo, which itself is thought to be driven by differential rotation in the stellar interior (e.g., Parker 1955), a phenomenon that has been confirmed in the Sun through helioseismology (Duvall et al. 1984). The observed decrease in X-ray emission from the pre-main sequence to the age of the Galactic field population can therefore be attributed to the rotational spin-down of the star, which is driven by mass loss through a magnetized stellar wind (Skumanich 1972).

The relationship between stellar rotation and tracers of magnetic activity is a vital probe of the stellar dynamo. A relationship between rotation and activity was first quantified by Pallavicini et al. (1981), who found that X-ray luminosity scaled as $L_X \propto (v \, \mathrm{sin} \, i)^{1.9}$, providing the first evidence for the dynamo-induced nature of stellar coronal activity. For very fast rotators the relationship was found to break down with X-ray luminosity reaching a saturation level of $L_X / L_{bol} \sim 10^{-3}$ (Micela et al. 1985), independent of spectral type. This saturation level is reached at a rotation period that increases toward later spectral types (Pizzolato et al. 2003), but it is unclear what causes this. Despite much work there is yet to be a satisfactory dynamo theory that can explain both the solar dynamo and that of rapidly rotating stars and the continued lack of a sufficiently large and unbiased sample has no doubt contributed to this.

We have produced a new catalog of stars with stellar rotation periods and X-ray luminosities, the details of which are presented in Wright et al. (2011). The catalog includes 824 solar- and late-type stars, including 445 field stars and 379 stars in nearby open clusters (ages 40--700~Myrs). The sample was homogenized by recalculating all X-ray luminosities and converting them all onto the ROSAT $0.1 - 2.4$~keV band. To minimize biases we removed all sources known to be X-ray variable, those that exhibit signs of accretion, or those in close binary systems. The sample is approximately equally distributed across the color range $V-K_s = 1.5 - 5.0$ (G2 to M4) with $\sim$30 stars per subtype, dropping to $\sim$10 stars per subtype outside of this, from F7 to M6.

Here we highlight some of the results derived from this sample, focussing attention on the regimes of low-activity dynamo decay and high-activity supersaturation. Details of these findings and further results and discussions from this sample can be found in Wright et al. (2011). Finally we present a new population synthesis model that can be used to address similar problems in stellar activity.

\begin{figure*}
\begin{center}
\includegraphics[height=450pt, angle=270]{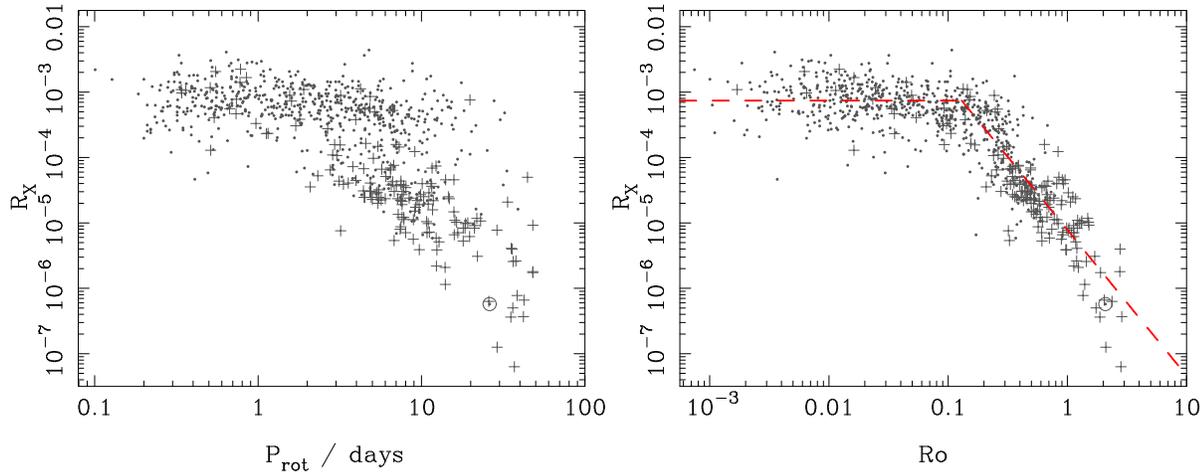}
\caption{$R_X = L_X / L_{bol}$ plotted against $P_{rot}$ (left) and the Rossby number, $Ro = P_{rot} / \tau$ (right), for all stars in our sample. Known binary stars are shown as plus symbols, and the Sun with a solar symbol. The best-fitting saturated and non-saturated activity--rotation relations are shown as a dashed red line in the right-hand panel.}
\label{allstars}
\end{center}
\end{figure*}

\section{Dynamo Efficiency in the Linear Regime}

The compiled catalog has been used to study the relationship between the level of X-ray activity, represented by the ratio of X-ray to bolometric luminosity, $R_X = L_X / L_{bol}$ and the rotation period, parameterized by the spectral-type independent Rossby number, $Ro = P_{rot} / \tau$, the ratio of the rotation period to the convective turnover time (Noyes et al. 1984). Figure~\ref{allstars} shows $R_X$ as a function of $P_{rot}$ and $Ro$, illustrating that the use of the latter parameter greatly reduces the scatter in the rotation--activity diagram, and highlights the two main regimes of coronal activity: a {\it linear} regime where activity increases with decreasing rotation period, and a {\it saturated} regime where the X-ray luminosity ratio is constant with log~$R_X = -3.13 \pm 0.08$. The transition between these two regimes is found to occur at $Ro = 0.13 \pm 0.02$ from a two-part power-law fit.

The sample used here suffers from a number of biases, most importantly an X-ray luminosity bias due to the selection only of sources with measured X-ray fluxes and photometric rotation periods. To overcome this we used an X-ray unbiased subset of our sample, the 36 Mt. Wilson stars with measured rotation periods (Donahue et al. 1996), all of which are detected in X-rays. While some biases may still exist due to the ability to measure rotation periods, Donahue et al. (1996) conclude that any such biases are unlikely to affect the rotation period distribution.

\begin{figure}
\begin{center}
\includegraphics[height=230pt, angle=270]{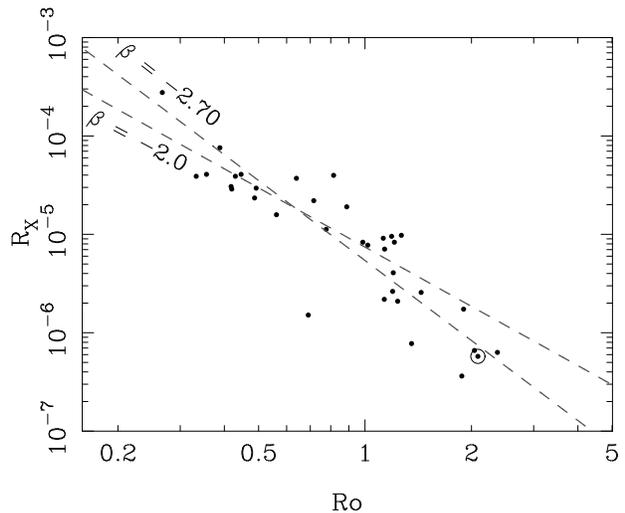}
\caption{$R_X = L_X / L_{bol}$ versus Rossby number for an unbiased sample of 36 stars with unsaturated X-ray emission. The log-log OLS bisector fit, $\beta = -2.70$, is shown as a dashed line alongside a fit with the canonical slope of $\beta = -2.0$.}
\label{unbiased}
\end{center}
\end{figure}

A single-part power-law fit in the linear regime (where $R_X \propto Ro^\beta$) to this sample is shown in Figure~\ref{unbiased} with a fit of $\beta= -2.70 \pm 0.13$. This is a steeper slope than fit using the full sample and notably steeper than the canonical value of $\beta = -2$ (Pallavicini et al. 1981), though their use of projected rotational velocities instead of rotation periods represents a different relationship than that fitted here. Our slope is inconsistent with the canonical value to a confidence of 5$\sigma$, which argues against a distributed dynamo operating throughout the convection zone, the efficiency of which scales as $Ro^{-2}$ (Noyes et al. 1984), and instead argues for an interface dynamo (e.g., Parker 1993), which has a more complex dependency where $\beta \neq -2$.

\section{Supersaturation}

\begin{figure*}
\begin{center}
\includegraphics[height=450pt, angle=270]{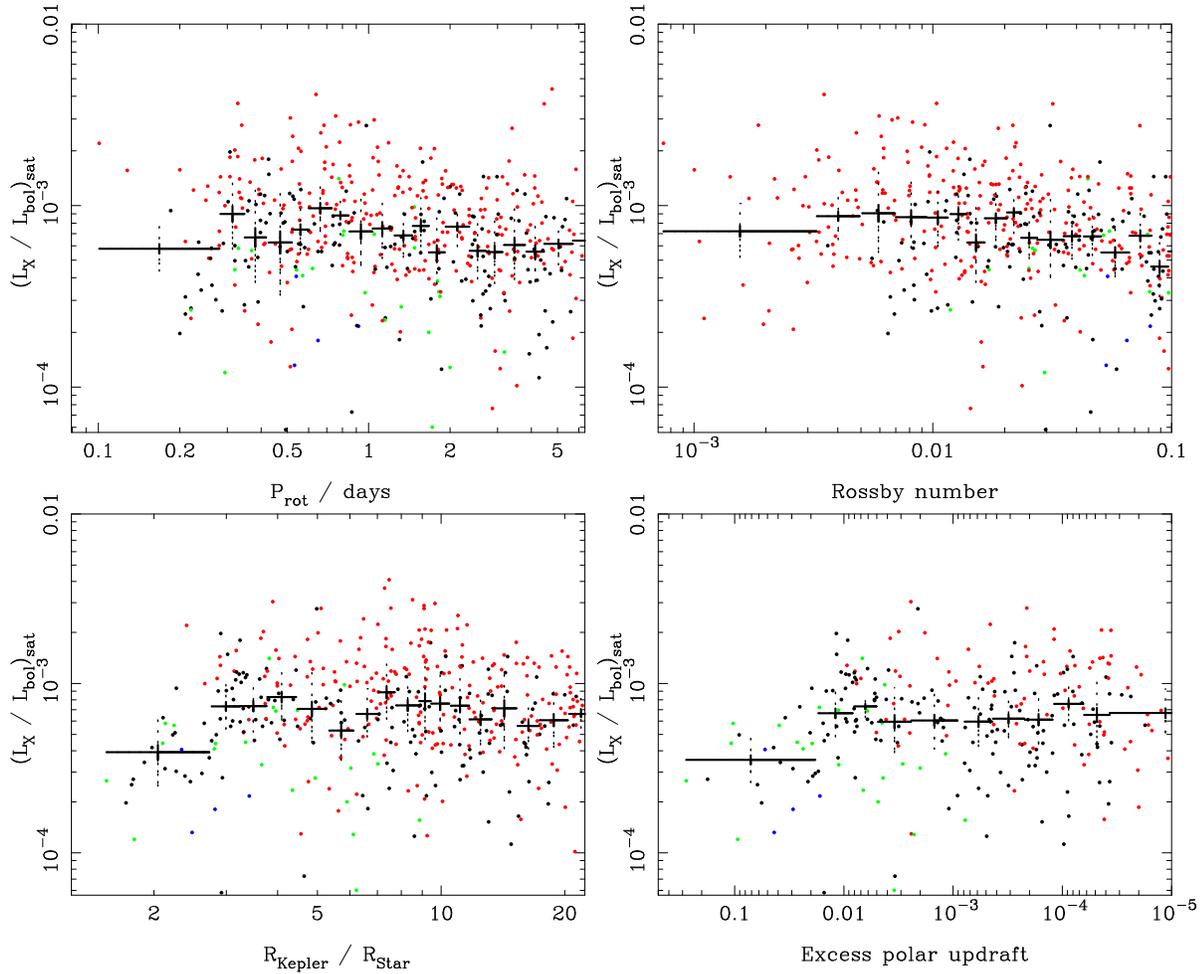}
\caption{$L_X / L_{bol}$ for stars with saturated X-ray emission as a function of the rotation period (upper left), the Rossby number (upper right), the Keplerian co-rotation radius (lower left), and the excess polar updraft (lower right). Stars are colored as per their spectral type: F-type (blue), G-type (green), K-type (black), and M-type (red). Also shown are 25-star average luminosity ratios, with standard errors.}
\label{supersaturation}
\end{center}
\end{figure*}

At very high rotational velocities the fractional X-ray luminosity is observed to decrease below the saturation level (Randich et al. 1996), an effect dubbed {\it supersaturation}, the cause of which is still heavily debated. Figure~\ref{supersaturation} shows $L_X / L_{bol}$ for all stars with saturated X-ray emission as a function of various quantities suggested as parameters of supersaturation. The first two panels show the effect of rotation period and Rossby number on the X-ray luminosity ratio, neither of which show evidence for supersaturation (i.e. a decline in $L_X / L_{bol}$) for either the fastest rotators or the sources with the smallest Rossby numbers.

Two theories suggested to explain the supersaturation effect are {\it coronal stripping} and {\it convective updrafts}. In the theory of coronal stripping the star is spinning so fast that centrifugal forces strip away the outer layers of the corona, reducing the X-ray emitting volume (Jardine \& Unruh 1999, James et al. 2000). This theory is parameterized by the corotation radius, $R_{Kepler}$, the radius at which centrifugal forces balance gravitational forces in the corona. In Figure~\ref{supersaturation} it is clear that this parameter exhibits a correlation with the supersaturation effect with the fastest rotators showing a drop to log~$R_X = -3.41 \pm 0.26$.

In the theory of convective updrafts, nonuniform heating of the convective envelope (in accordance with the von Zeipel, 1924, theorem for rapidly rotating stars) results in a poleward migration of active regions and a reduction in the filling factor of active regions on the stellar surface (St\c epie\'n et al. 2001). This can be quantified in terms of the excess polar updraft at the bottom of the convective envelope and is also shown in Figure~\ref{supersaturation}, where the parameter also displays a strong correlation with the supersaturation effect with the fastest rotators dropping to log~$R_X = -3.45 \pm 0.16$.

Based on this data it is difficult to distinguish between these two theories, partly because they have similar dependencies in terms of stellar parameters. An alternative distinguishing method is the underlying coronal structure that the two theories require to be effective. Coronal stripping will occur if the loops are large, low density, and unstable to the centrifugal force. In Figure~\ref{supersaturation} a decrease in X-ray luminosity begins at $R_{Kepler} / R_\star \sim 3$, suggesting that coronal loops are as large as two stellar radii in height. Alternatively coronal loops may be compact, high density and influenced by the dynamics of surface flows where convective updrafts may influence the surface activity level. The difference between the two supersaturation theories is therefore the size and density of the coronal plasma. High-resolution X-ray spectroscopy can be used to infer coronal plasma densities (and therefore scale heights), but such data for supersaturated stars is rare. The only such observation to date is of VW~Cephei, for which coronal loop heights of 0.06-0.2~$R_\star$ have been measured (Huenemoerder et al. 2006). The rotational period of VW~Cephei implies $R_{Kepler} / R_\star \simeq 1.75$, significantly higher than the observed coronal loop heights, and suggesting that coronal stripping is unlikely to be responsible for supersaturation.



\section{Modeling the Decay of Stellar Dynamos}

\begin{figure}
\begin{center}
\includegraphics[height=210pt, angle=270]{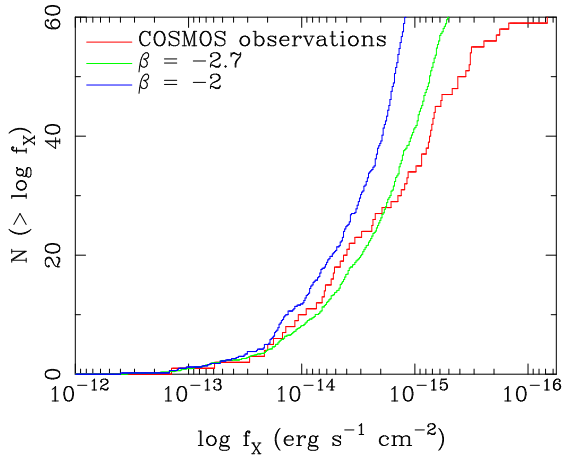}
\caption{Cumulative distribution of X-ray fluxes for sources in the COSMOS field compared to two predictions from the XStar model for $\beta = -2.0$ and $\beta = -2.7$. The observations are complete to $f_X \sim 2 \times 10^{-15}$~erg~s$^{-1}$~cm$^{-2}$.}
\label{xstar}
\end{center}
\end{figure}

Studies of the rotation -- activity relation in the manner performed above are limited both by the sample size of stars with measured X-ray luminosities and rotation periods and also by biases inherent in compiling such samples. A different approach that allows one to overcome such problems is to model the X-ray emission of a population of stars and compare the number and properties of such stars with observations. We have compiled such a model, XStar, combining a Galactic population synthesis model with models of rotational spin-down (e.g., Barnes 2003) and the activity--rotation framework outlined here. The results of this model can then be compared to X-ray observations that sample different sight-lines in our Galaxy such as deep targeted observations or wide-field, shallow surveys.

An example of the model is shown in Figure~\ref{xstar} where two realizations of the model using $\beta = -2.0$ and $\beta = -2.7$ are compared to the cumulative flux distribution of X-ray sources in the high Galactic latitude {\it Chandra} COSMOS field (Wright, Drake \& Civano 2010). The models reproduce the observations very well down to the estimated observational completeness limit of $f_X \sim 2 \times 10^{-15}$~erg s$^{-1}$ cm$^{-2}$, with similar forms and the brightest sources reproduced well. The $\beta = -2.7$ model provides the better fit, although various other parameters within the model must be fully tested before firm conclusions can be drawn. Nonetheless the code does will in reproducing the properties (distances, spectral types, X-ray fluxes) of the observations, and will therefore prove useful for studying the properties of mature, X-ray emitting stellar populations as well as for estimating stellar contamination rates for studies of star forming regions and extra-Galactic populations (e.g., Brassington et al. 2012).

\section{Summary}

A new catalog of stars with measured X-ray luminosities and rotation periods is used to study the rotation -- activity relation. The power-law slope of the unsaturated regime is fit as $\beta = -2.7 \pm 0.13$, inconsistent with the canonical $\beta = -2$ value with a confidence of 5$\sigma$ and arguing for an interface-type dynamo. Coronal supersaturation in ultra-fast rotators is shown to correlate well with both the corotation radius and the excess polar updraft as the parameters of the coronal stripping and convective updrafts theories, though other observations support the latter theory. Finally a new population synthesis model of X-ray emitting stars in a Galactic sightline is introduced and tested successfully on deep observations of the high Galactic latitude COSMOS field. A better fit is found for $\beta = -2.7$ than for $\beta = -2$, confirming results with a fixed sample.


\begin{thebibliography}{}

\bibitem[{{Barnes}(2003)}]{barn03}
{Barnes}, S.~A. 2003, \apj, 586, 464

\bibitem[{{Brassington} {et~al.}(2012){Brassington}, {Fabbiano}, {Zezas}, {Kundu}, {Kim}, {Fragos}, {King}, {Pellegrini}, {Trinchieri}, {Zepf}, \& {Wright}}]{bras12}
{Brassington}, N.~J., {et~al.} 2012, \apj, 755, 162

\bibitem[{{Donahue} {et~al.}(1996){Donahue}, {Saar}, \& {Baliunas}}]{dona96}
{Donahue}, R.~A., {Saar}, S.~H., \& {Baliunas}, S.~L. 1996, \apj, 466, 384

\bibitem[{{Duvall} {et~al.}(1984){Duvall}, {Dziembowski}, {Goode}, {Gough},  {Harvey}, \& {Leibacher}}]{duva84}
{Duvall}, Jr., T.~L., {Dziembowski}, W.~A., {Goode}, P.~R., {Gough}, D.~O., {Harvey}, J.~W., \& {Leibacher}, J.~W. 1984, Nature, 310, 22
  
\bibitem[{{Huenemoerder} {et~al.}(2006){Huenemoerder}, {Testa}, \&  {Buzasi}}]{huen06}
{Huenemoerder}, D.~P., {Testa}, P., \& {Buzasi}, D.~L. 2006, \apj, 650, 1119

\bibitem[{{Jardine} \& {Unruh}(1999)}]{jard99}
{Jardine}, M., \& {Unruh}, Y.~C. 1999, Astron. \& Astrop., 346, 883

\bibitem[{{James} {et~al.}(2000){James}, {Jardine}, {Jeffries}, {Randich},  {Collier Cameron}, \& {Ferreira}}]{jame00}
{James}, D.~J., {Jardine}, M.~M., {Jeffries}, R.~D., {Randich}, S., {Collier Cameron}, A., \& {Ferreira}, M. 2000, \mnras, 318, 1217

\bibitem[{{Micela} {et~al.}(1985){Micela}, {Sciortino}, {Serio}, {Vaiana}, {Bookbinder}, {Golub}, {Harnden}, \& {Rosner}}]{mice85}
{Micela}, G., {et~al.}, 1985, \apj, 292, 172

\bibitem[{{Noyes} {et~al.}(1984){Noyes}, {Hartmann}, {Baliunas}, {Duncan}, \&  {Vaughan}}]{noye84}
{Noyes}, R.~W., {Hartmann}, L.~W., {Baliunas}, S.~L., {Duncan}, D.~K., \&  {Vaughan}, A.~H. 1984, \apj, 279, 763

\bibitem[{{Pallavicini} {et~al.}(1981){Pallavicini}, {Golub}, {Rosner}, {Vaiana}, {Ayres}, \& {Linsky}}]{pall81}
{Pallavicini}, R., {Golub}, L., {Rosner}, R., {Vaiana}, G.~S., {Ayres}, T., \& {Linsky}, J.~L. 1981, \apj, 248, 279

\bibitem[{{Parker}(1955)}]{park55}
{Parker}, E.~N. 1955, \apj, 122, 293

\bibitem[{{Parker}(1993)}]{park93}
---. 1993, \apj, 408, 707

\bibitem[{{Pizzolato} {et~al.}(2003){Pizzolato}, {Maggio}, {Micela}, {Sciortino}, \& {Ventura}}]{pizz03}
{Pizzolato}, N., {Maggio}, A., {Micela}, G., {Sciortino}, S., \& {Ventura}, P. 2003, Astron. \& Astrop., 397, 147
  
\bibitem[{{Randich} {et~al.}(1996){Randich}, {Schmitt}, {Prosser}, \&  {Stauffer}}]{rand96}
{Randich}, S., {Schmitt}, J.~H.~M.~M., {Prosser}, C.~F., \& {Stauffer}, J.~R.  1996, Astron. \& Astrop., 305, 785

\bibitem[{{Skumanich}(1972)}]{skum72}
{Skumanich}, A. 1972, \apj, 171, 565

\bibitem[{{St{\c e}pie{\'n}} {et~al.}(2001){St{\c e}pie{\'n}}, {Schmitt}, \&  {Voges}}]{step01}
{St{\c e}pie{\'n}}, K., {Schmitt}, J.~H.~M.~M., \& {Voges}, W. 2001, Astron. \& Astrop., 370,  157

\bibitem[{{Telleschi} {et~al.}(2007){Telleschi}, {G{\"u}del}, {Briggs},  {Audard}, \& {Palla}}]{tell07}
{Telleschi}, A., {G{\"u}del}, M., {Briggs}, K.~R., {Audard}, M., \& {Palla}, F.  2007, Astron. \& Astrop., 468, 425

\bibitem[{{Vaiana} {et~al.}(1981){Vaiana}, {Cassinelli}, {Fabbiano},  {Giacconi}, {Golub}, {Gorenstein}, {Haisch}, {Harnden}, {Johnson}, {Linsky},  {Maxson}, {Mewe}, {Rosner}, {Seward}, {Topka}, \& {Zwaan}}]{vaia81}
{Vaiana}, G.~S., {et al.}, 1981, \apj, 245, 163

\bibitem[{{von Zeipel} (1924)}]{vonz24}
{von Zeipel}, H., 1924, MNRAS, 84, 665

\bibitem[{{Wright} {et~al.}(2010){Wright}, {Drake}, \& {Civano}}]{wrig10b}
{Wright}, N.~J., {Drake}, J.~J., \& {Civano}, F. 2010, \apj, 725, 480

\bibitem[{{Wright} {et~al.}(2011){Wright}, {Drake}, {Mamajek}, \& {Henry}}]{wrig12b}
{Wright}, N.~J., {Drake}, J.~J., {Mamajek}, E.~E., \& {Henry}, G.~W., 2012, \apj, 743, 48

\end{thebibliography}
\end{document}